\documentclass[preprint,prd,showpacs]{revtex4}
\usepackage{amsmath,epsfig}
\usepackage{amsfonts}
\usepackage{amssymb}
\usepackage{graphicx}
\usepackage[cmtip,arrow]{xy}
\usepackage{amssymb,amsopn}
\begin{document}

\title{Quantum string cosmology in the phase space}


\author{Rub\'en Cordero}\email{cordero@esfm.ipn.mx}
\affiliation{Departamento de F\'{\i}sica, Escuela Superior de
F\'{\i}sica y Matem\'aticas del IPN\\ Unidad Adolfo
L\'opez Mateos, Edificio 9, 07738, M\'exico D.F., M\'exico}

\author{Erik D\'{\i}az}\email{erik_harry@hotmail.com}
\affiliation{Departamento de F\'{\i}sica, Escuela Superior de
F\'{\i}sica y Matem\'aticas del IPN\\ Unidad Adolfo
L\'opez Mateos, Edificio 9, 07738, M\'exico D.F., M\'exico}

\author{Hugo Garc\'{\i}a-Compe\'an}
\email{compean@fis.cinvestav.mx} \affiliation{Departamento
de F\'{\i}sica, Centro de Investigaci\'on y de Estudios
Avanzados del IPN\\ P.O. Box 14-740, 07000 M\'exico D.F., M\'exico}

\author{Francisco J. Turrubiates}\email{fturrub@esfm.ipn.mx}
\affiliation{Departamento de F\'{\i}sica, Escuela Superior de F\'{\i}sica
y Matem\'aticas del IPN\\ Unidad Adolfo
L\'opez Mateos, Edificio 9, 07738, M\'exico D.F., M\'exico.}
\date{\today}

\begin{abstract}
Deformation quantization is applied to quantize gravitational
systems coupled with matter. This quantization procedure is
performed explicitly for quantum cosmology of these systems in a
flat minisuper(phase)space. The procedure is employed in a
quantum string minisuperspace corresponding to an axion-dilaton
system in an isotropic FRW Universe. The Wheeler-DeWitt-Moyal
equation is obtained and its corresponding Wigner function is given
analytically in terms of Meijer's functions. Finally, this Wigner
functions is used to extract physical information of the system.
\end{abstract}

\maketitle

%%%%%%%%%%%%%%%%%%%%%%%%%%%%%%%%%%%%%%%%%%%%
%% MAINMATTER
%%%%%%%%%%%%%%%%%%%%%%%%%%%%%%%%%%%%%%%%%%%%

\section{Introduction}
Our contribution to the {\it VIII Mexican workshop on gravitation and mathematical-physics 2010} focused on explaining the implementation of the deformation quantization formalism to a gravitational field coupled to matter. Also it was considered its application to diverse cosmological models and the baby Universe system. The analysis included a minisuperspace approach of the de Sitter model, the Kantowski-Sachs Universe (for the commutative and non-commutative cases) and finally we discussed the case of string cosmology with a dilaton exponential potential.  The detailed construction is presented in \cite{Cordero:2011xa} (and references therein)
and will not be repeated here. Instead we will apply the formalism to another case using the minisuperspace approach of string cosmological models involving axion and dilaton fields in a curved space.
The issues we will discuss in section two involve mainly a brief review of the Weyl-Wigner-Groenewold-Moyal (WWGM) formalism of deformation quantization \cite{Zachos} in the context of quantum cosmological models. This is an integral functional formalism and it is well defined in the case of the whole Wheeler's superspace of 3-metrics in the ADM's description of gravity and matter.

The four dimensional low energy effective field theory action of string theory contains at least three massless fields: the graviton, the axion and the dilaton \cite{Polchinski}. Indirect evidences of low energy string theory can appear in the physical consequences of the axion in a curved spacetime \cite{Saha}. In string theory the dilaton is quite relevant since it defines the string coupling constant.

One of the most important models of string cosmology is the pre-big-bang scenario \cite{Gasperini:2002bn}. It incorporates the target space string duality through the {\it scale factor duality}. This predicts a decreasing curvature for negative values of the time coordinate. This curvature is the specular image of the so-called {\it post-big-bang} cosmology. The pre-big-bang cosmology scenario is important as it is a modification produced by string theory which is (at least perturbatively) a consistent theory of quantum gravity. Thus it is expected to describe the correct modification to standard general relativity at very early times. The prediction is that there is not a standard big-bang singularity since it is smoothed by the scale factor duality coming from target space duality. In the present paper we study precisely the quantum cosmological model regarding the pre-big-bang scenario in the context of the quantization of the phase space of the minisuperspace of these models. In particular we discuss the cosmological model of a string theory in four dimensions with axion and dilaton fields in a curved space \cite{Dabrowski}.

The rest of the paper is organized as follows. In section three, we apply the deformation quantization formalism to string cosmology with axion and dilation in a curved minisuperspace approach and we find the exact Wigner functional which is equivalent to the quantum states of the Universe. In the last section, we give our final remarks.

%%%%%%%%%%%%%%%%%%%%%%%%%%%%%%%%%%%%%%%%%%%%%%%%%%%%%%
%%%%%%%%%%%%%%%%%%%%%%%%%%%%%%%%%%%%%%%%%%%%%%%%%%%%%%
\section{Deformation quantization of the gravitational field coupled to matter}

We start by giving a brief overview of the Hamiltonian formalism for the gravitational field. In particular we consider the ADM decomposition of general relativity coupled to matter (see for instance \cite{IntQC}). We take a globally hyperbolic spacetime modeled with a pseudo Riemannian manifold $(M,g)$ where $M= \Sigma \times {R}$ and $ds^2=g_{\mu \nu} dx^\mu dx^\nu
= - (N^2-N^iN_i) (dt)^2 + 2N_idx^idt + h_{ij} dx^i dx^j,$
with signature $(-,+,+,+)$. Here $h_{ij}$ is the intrinsic metric on the hypersurface $\Sigma$, $N$ is the lapse function and $N^{i}$ is the shift vector.
The {\it superspace} is defined by ${\tt Riem}(\Sigma)=\{h_{ij}(x), \  \Phi(x) | x \in \Sigma \}$ where $\Phi$ is the scalar field. Let ${\tt Met}(\Sigma)=\{ h_{ij}(x) |
x \in \Sigma\}$ which is an infinite dimensional manifold. The moduli space of the theory is ${\cal M}={\frac{{\tt Riem}(\Sigma)}{{\tt Diff}(\Sigma)}}$ or for pure gravity ${\cal M}={\frac{{\tt
Met}(\Sigma)}{{\tt Diff}(\Sigma)}}$ where ${\tt Diff}(\Sigma)$ is the group of diffeomorphisms of $\Sigma$. The corresponding phase space is given by $\Gamma^*\cong T^* {\tt Met}(\Sigma)= \{(h_{ij}(x), \pi^{ij}(x))\}$,
where $\pi^{ij}=\frac{\partial{L}_{EH}}{\partial \dot{h}_{ij}}$ and
${L}_{EH}$ is the Einstein-Hilbert Lagrangian. In the following we
will deal with fields at the moment $t=0$ (on $\Sigma$) and we put
$h_{ij}(x,0) \equiv h_{ij}(x)$ and $\pi^{ij}(x,0) \equiv \pi^{ij}(x)$.

The kind of systems considered here are invariant under diffeomorphisms thus their canonical Hamiltonian is pure constraint. Therefore we only should worry to solve the constraints of the theory. These are the momentum and the Hamiltonian constraint. This latter is written as
\begin{equation}
{\cal H}_\perp(x) =4 {\vartheta}^2 G_{ijkl}\pi^{ij} \pi^{kl} - \frac{\sqrt{h}}{4 \vartheta^2} \big({^3R} - 2 \Lambda\big) + \frac{1}{2} \sqrt{h}
\bigg(\frac{\pi^2}{h} + h^{ij} \Phi_{,i} \Phi_{,j} + 2 V(\Phi) \bigg) =0,
\label{eqn:constraints1}
\end{equation}
where $\vartheta^{2} = 4\pi G_{N}$, $\pi=\frac{\partial L_M}{\partial \dot{\Phi}}$, with $L_M$ stands for the matter Lagrangian, $\Lambda$ is the cosmological constant, ${^3R}(h)$ is the scalar curvature of $\Sigma$, $G_{ijkl} = \frac{1}{2}h^{-1/2} \big(h_{ik}h_{jl} + h_{il} h_{jk} - h_{ij}h_{kl} \big)$, ${}_{,j}$ denotes partial derivative with respect to $x^{j}$ and $h$ is the determinant of $h_{ij}$. One of the most important structures for quantization is the Poisson bracket between $h_{ij}$ and $\pi^{kl}$ given by
\begin{equation}
\{h_{ij}(x), \pi^{kl}(y) \}_{PB}= \frac{1}{2} (\delta^k_i \delta^l_j
+ \delta^k_j \delta^l_i) \delta(x-y).
\end{equation}
The canonical quantization promotes the canonical variables to operators acting on some Hilbert space (or Fock space). In the $h$-representation they look like:
$\widehat{h}_{ij} |h_{ij},\Phi \rangle =h_{ij} |h_{ij},\Phi \rangle$,
$\widehat{\pi}^{ij} |h_{ij},\Phi \rangle =-i \hbar
\frac{\delta}{\delta h_{ij}(x)} |h_{ij},\Phi \rangle$,
$\widehat{\Phi} |h_{ij}, \Phi \rangle =\Phi(x) |h_{ij},\Phi \rangle$
and $\widehat{\pi}_\Phi |h_{ij},\Phi \rangle =-i \hbar
\frac{{\delta}}{\delta \Phi(x)} |h_{ij},\Phi \rangle.$ These
operators satisfy the commutation relations
\begin{equation}
[\widehat{h}_{ij}(x), \widehat{\pi}^{kl}(y)] = \frac{i \hbar}{2} (\delta^k_i \delta^l_j + \delta^k_j \delta^l_i)
\delta(x-y),
\end{equation}
and similar expressions for the matter part. The Hamiltonian constraint at the quantum level is given by $\widehat{\cal H}_\perp | \Psi \rangle =0.$ In the $h$-representation we have the constraint given by
\begin{equation}
\bigg[-4 \vartheta^2G_{ijkl} \frac{\delta^2}{\delta h_{ij} \delta
h_{kl}} + \frac{\sqrt{h}}{4 \vartheta^2} \bigg( - \ {^3R}(h) + 2
\Lambda
+ 4 \kappa^2 \widehat{T}^{00} \bigg) \bigg]
\Psi[h_{ij},\Phi] =0 \,\, ,
\end{equation}
where $\langle h_{ij},\Phi | \Psi \rangle =  \Psi[h_{ij},\Phi]$ and $\widehat{T}^{00}=-\frac{1}{2 h} \frac{\delta^2}{\delta \Phi^2} + \frac{1}{2} h^{ij} \Phi_{,i} \Phi_{,j} + V(\Phi)$. This equation is called the Wheeler-DeWitt (WDW)
equation for the wave function of the Universe $\Psi[h_{ij},\Phi]$.

%%%%%%%%%%%%%%%%%%%%%%%%%%%%%%%%%%%%%%%%%%%%%%%%%%%%%%%%%
The deformation quantization of gravity in ADM formalism and constrained systems has been considered in \cite{Antonsen:1997yq}. In the rest of the paper we assume that the superspace has a flat metric
$G_{ijkl}$ to ensure the existence of the Fourier transform.

Let $F[h_{ij},\pi^{ij}; \Phi,\pi_{\Phi}]$ be a functional on the
phase space $\Gamma^*$ (Wheeler's phase superspace) and let
$\widetilde{F}[\mu^{ij},\lambda_{ij};\mu,\lambda]$
be its Fourier transform
$$
\widetilde{F}[\mu^{ij},\lambda_{ij};\mu,\lambda] = \int  {\cal D} \pi^{ij} {\cal D}h_{ij} {\cal D} \pi_{\Phi} {\cal D}\Phi  \exp\bigg\{-i \int dx \bigg(  \mu^{ij}({x}) h_{ij}({x}) + \lambda_{ij}({x})
 \pi^{ij}({x})
$$
\begin{equation}
+  \mu({x}) \Phi({x}) +  \lambda({x})
 \pi_{\Phi}({x}) \bigg) \bigg\}
F[h_{ij},\pi^{ij};\Phi,\pi_{\Phi}],
\end{equation}
where  ${\cal D}h_{ij} = \prod_{x} d h_{ij}(x)$, ${\cal D} \pi^{ij}=\prod_{x} d \pi^{ij}(x)$, ${\cal D}\Phi =
\prod_{x} d \Phi (x)$, ${\cal D} \pi_{\Phi} = \prod_{x} d \pi_{\Phi}(x)$. By analogy to the quantum mechanics case, we define the Weyl quantization map ${\cal W}$ as follows
\footnotesize
\begin{equation}
\widehat{F}= {\cal W}(F[h_{ij},\pi^{ij}; \Phi,\pi_{\Phi}])
:= \int {\cal D} \left(\frac{\lambda_{ij}}{2 \pi}\right) {\cal
D}\left(\frac{\mu^{ij}}{2 \pi}\right) {\cal D} \left(\frac{\lambda}{2 \pi}\right) {\cal D}\left(\frac{\mu}{2 \pi}\right)
\widetilde{F}[\mu^{ij},\lambda_{ij};\mu,\lambda] \widehat{\cal U}[\mu^{ij},\lambda_{ij};\mu,\lambda],
\end{equation}
\normalsize
where
\footnotesize
\begin{equation}
\widehat{\cal U}[\mu^{ij},\lambda_{ij};\mu,\lambda]:= \exp \bigg\{i\int dx \bigg( \mu^{ij}({x}) \widehat{h}_{ij}({x}) + \lambda_{ij}({x})
\widehat{\pi}^{ij}({x})+ \mu({x})
\widehat{\Phi}({x}) + \lambda({x}) \widehat{\pi}_{\Phi}({x})  \bigg)\bigg\}.
\end{equation}
\normalsize
Here $\widehat{h}_{ij}$, $\widehat{\pi}^{ij}$, $\widehat{\Phi}$ and $\widehat{\pi}_{\Phi}$ are the field operators defined by: $\widehat{h}_{ij}({x}) |h_{ij},\Phi \rangle = h_{ij}({x})
|h_{ij},\Phi \rangle$, $\widehat{\pi}_{ij}({x})
|\pi_{ij},\pi_{\Phi} \rangle = \pi_{ij}({x}) |\pi_{ij},\pi_{\Phi}\rangle,$ $\widehat{\Phi}({x}) |h_{ij},\Phi \rangle = \Phi(x)|h_{ij},\Phi \rangle,$ $ \widehat{\pi}_{\Phi}({x})
| \pi_{ij}, \pi_{\Phi} \rangle = \pi_{\Phi}({x}) |\pi_{ij}, \pi_{\Phi} \rangle \,.$
The Campbell-Baker-Hausdorff formula, commutator algebra and the completeness relations lead to an explicit form for the operator $\widehat{\cal U}$ to be
$$
\widehat{\cal U}[\mu^{ij},\lambda_{ij};\mu,\lambda]
=
\int {\cal D} {h}_{ij} {\cal D} {\Phi} \exp \bigg \{i \int dx \mu^{ij}({x}) h_{ij}({x})  + \mu({x}) \Phi({x})\bigg \}
$$
\begin{equation}
 \times \bigg| h_{ij} - \frac{\hbar \lambda_{ij}}{2}, \Phi - \frac{\hbar \lambda}{2} \bigg\rangle
\bigg\langle h_{ij} + \frac{\hbar \lambda_{ij}}{2}, \Phi + \frac{\hbar \lambda}{2}\bigg|.
\end{equation}
One immediately obtains the following structure
\begin{equation}
\widehat{F} = \int {\cal D} \left(\frac{\pi^{ij}}{2 \pi
\hbar}\right) {\cal D} h_{ij}  {\cal D} \left(\frac{\pi_{\Phi}}{2
\pi \hbar}\right) {\cal D} \Phi
F[h_{ij},\pi^{ij}; \Phi,\pi_{\Phi}]
\widehat{\Omega} [h_{ij},\pi^{ij}; \Phi,\pi_{\Phi}],
\end{equation}
where the operator $\widehat{\Omega}$ is the Stratonovich-Weyl quantizer and it is given by
\footnotesize
$$
\widehat{\Omega} [h_{ij},\pi^{ij}; \Phi,\pi_{\Phi}] = \int {\cal D} \left(\frac{\hbar \lambda_{ij}}{ 2 \pi}\right)  {\cal D}
\mu^{ij} {\cal D} \left(\frac{\hbar \lambda}{2 \pi}\right)  {\cal D} \mu
$$
\begin{equation}
\times \exp \bigg \{ -i \int dx
\bigg(\mu^{ij}(x) h_{ij} (x) + \lambda_{ij}(x)\pi^{ij} (x) + \mu(x) \Phi(x)
+ \lambda(x) \pi_{\Phi}(x) \bigg) \bigg \}
\widehat{\cal U} [\mu^{ij},\lambda_{ij};\mu,\lambda].
\end{equation}
\normalsize
This operator can be written in the following form (that can be very useful to invert the mapping ${\cal W}$)
\footnotesize
$$
\widehat{\Omega}[h_{ij},\pi^{ij}; \Phi,\pi_{\Phi}] =\int {\cal D} \xi_{ij}  \int {\cal D} \xi\exp \bigg \{ - \frac{i}{\hbar} \int dx \xi_{ij}(x) \pi^{ij}(x) + \xi(x)
\pi_{\Phi}(x) \bigg \}
$$
\begin{equation}
\times \bigg|h_{ij} - \frac{\xi_{ij}}{2}, \Phi - \frac{\xi}{2} \bigg\rangle \bigg\langle h_{ij} + \frac{\xi_{ij}}{2},
\Phi + \frac{\xi}{2} \bigg|.
\end{equation}
\normalsize
The space ${\cal A}$ of all functionals on the phase space $\Gamma^*$ i.e. ${\cal A}:= \{F=F[h_{ij},\pi^{ij}; \Phi,\pi_{\Phi}]\}$, forms with the usual product an associative and commutative algebra. This algebra can be deformed into an associative and non-commutative algebra ${\cal A}^\star$ with the $\star-$product. In relation to the ${\cal W}$ map this $\star$ product is defined as: ${\cal W}^{-1}(\widehat{F}\cdot \widehat{G}) =F \star G$ for any pair of functionals $F$ and $G$ in ${\cal A}$.
Thus the $\star$ product is defined as
\begin{equation}
(F\star G)[h_{ij},\pi^{ij}; \Phi,\pi_{\Phi}]:= {\cal W}^{-1}(\widehat{F} \widehat{G})
= {\rm Tr} \bigg \{
\widehat{\Omega}[h_{ij},\pi^{ij}; \Phi,\pi_{\Phi}] \widehat{F} \widehat{G} \bigg \},
\end{equation}
or after some straightforward computations
\begin{equation}
\big(F \star  G\big)[h_{ij},\pi^{ij}; \Phi,\pi_{\Phi}]
 = F[h_{ij},\pi^{ij}; \Phi,\pi_{\Phi}] \exp\bigg\{\frac{i\hbar}{2}
\buildrel{\leftrightarrow}\over {\cal P}\bigg\} G[h_{ij},\pi^{ij}; \Phi,\pi_{\Phi}],
\end{equation}
where $\buildrel{\leftrightarrow}\over {\cal P}$ is the operator given by
\footnotesize
\begin{equation}
\buildrel{\leftrightarrow}\over {\cal P} := \int dx \bigg(\frac{{\buildrel{\leftarrow}\over {\delta}}}{\delta h_{ij}(x)}
\frac{{\buildrel{\rightarrow}\over {\delta}}}{\delta \pi^{ij}(x)} - \frac{{\buildrel{\leftarrow}\over {\delta}}}{\delta
\pi^{ij}(x)} \frac{{\buildrel{\rightarrow}\over{\delta}}}{\delta h_{ij}(x)}\bigg)
 +  \int dx
\bigg(\frac{{\buildrel{\leftarrow}\over {\delta}}}{\delta \Phi(x)} \frac{{\buildrel{\rightarrow}\over {\delta}}}{\delta
\pi_{\Phi}(x)} - \frac{{\buildrel{\leftarrow}\over {\delta}}}{\delta \pi_{\Phi}(x)} \frac{{\buildrel{\rightarrow}\over
{\delta}}}{\delta \Phi(x)}\bigg).
\end{equation}
\normalsize
Let $\widehat{\rho}$ be the density operator of a quantum state.
The functional $\rho_{W}[h_{ij},\pi^{ij}; \Phi,\pi_{\Phi}]$ corresponding to $\widehat{\rho}$ reads
$$
\rho_{W}[h_{ij},\pi^{ij}; \Phi,\pi_{\Phi}] = {\rm Tr} \bigg \{ \widehat{\Omega}[h_{ij},\pi^{ij}; \Phi,\pi_{\Phi}]
\widehat{\rho}\bigg \} \nonumber
$$
$$
: = \int {\cal D} \left(\frac{\xi_{ij}}{2 \pi \hbar}\right)  {\cal D} \left(\frac{\xi}{2 \pi
\hbar}\right) \exp \bigg\{ - \frac{ i}{\hbar} \int dx \ \bigg( \xi_{ij}(x) \pi^{ij}(x) +  \xi(x) \pi_{\Phi}(x) \bigg) \bigg\}
$$
\begin{equation}
\times \bigg\langle h_{ij} + \frac{\xi_{ij}}{2}, \Phi + \frac{\xi}{2} \bigg| \widehat{\rho} \bigg| h_{ij} - \frac{\xi_{ij}}{2}, \Phi - \frac{\xi}{2} \bigg\rangle.
\end{equation}
For a pure state of the system $\widehat{\rho} = |\Psi\rangle \langle \Psi |$ we have
\footnotesize
$$
\rho_{_W}[h_{ij},\pi^{ij}; \Phi,\pi_{\Phi}] = \int {\cal D} \left(\frac{\xi_{ij}}{2 \pi \hbar}\right){\cal D} \left(\frac{\xi}{2 \pi
\hbar}\right) \exp \bigg\{ - \frac{i}{\hbar} \int dx \ \bigg( \xi_{ij}(x) \pi^{ij}(x) + \xi(x) \pi_{\Phi}(x) \bigg) \bigg\}
$$
\begin{equation}
\times \Psi^* \left[h_{ij} - \frac{\xi_{ij}}{2}, \Phi - \frac{\xi}{2} \right] \Psi \left[h_{ij} + \frac{\xi_{ij}}{2}, \Phi +
\frac{\xi}{2}\right].
\end{equation}
\normalsize
It is now possible to write the Hamiltonian constraint
in terms of the $\star -$product and the Wigner functional as
\begin{equation}
{\cal H}_{\perp} \star \rho_{_W}[h_{ij},\pi^{ij}; \Phi,\pi_{\Phi}] = 0.
\label{MWDW}
\end{equation}
This is the Moyal deformation of the Wheeler-DeWitt equation an we will just called it the Wheeler-DeWitt-Moyal (WDWM) equation.

\section{Deformation Quantization in quantum cosmology}

String theory can be employed to describe the evolution of the early Universe and one of the most important areas of research are the cosmological consequences of the dilaton and its role in the pre-big-bang scenario \cite{Gasperini:2002bn}.
An interesting case to deal with under the deformation quantization procedure is the axion-dilaton quantum cosmology in curved space \cite{Dabrowski}.

\subsection{Axion-dilaton quantum cosmology in curved space}

Let's start with a model in a FRW metric with $\Lambda=0$, axion
energy density, dilaton field and spatial curvature described by the
following effective action
\begin{equation}
S=\frac{\lambda_{s}}{2} \int d\tau \left[-\overline{\phi}'^{2} + \beta'^{2}- e^{-2\overline{\phi}} \left(\frac{1}{2}q^{2} e^{-2\sqrt{3}\beta} - 6 \kappa e^{-2\beta /\sqrt{3}} \right) \right],
\end{equation}
where $\lambda_{s}$ denotes the string length, $q^{2}$ codifies the contribution of axion energy density \cite{Dabrowski}, $\kappa$ is related to the spatial curvature and the prime variables stand for the derivatives with respect to the \textit{dilaton time} \cite{Meissner}. \\
Using now the variables $\phi = \overline{\phi} + \sqrt{3}\beta, \  y =\overline{\phi} \sqrt{3}  + \beta = \sqrt{3}\left( \phi - 2\ln{a} -\ln \int\frac{d^{3}x}{\lambda_{s}^{3}} \right)$,
we obtain that
\begin{equation}
S=\frac{\lambda_{s}}{4} \int d\tau \left[{\phi}'^{2} - y'^{2}- q^{2}e^{-2\phi} + 12 \kappa e^{-2y /\sqrt{3}}  \right].
\end{equation}
The corresponding WDW equation takes the following form
\begin{equation}
\widehat{H}\Psi(y,\phi)=\frac{1}{\lambda_s}\left(\hbar^{2} \partial_y^2- \hbar^{2} \partial_\phi^2+\frac{1}{4}\lambda_s^2q^2e^{-2\phi}-3\lambda_s^2\kappa e^{-\frac{2}{\sqrt{3}}y}\right)\Psi(y,\phi)=0.
\end{equation}
The solutions of this equation are obtained by the method of separation of variables $\Psi(y,\phi)=\chi_\alpha(y)\psi_\alpha(\phi)$.
In this way we get for the $\chi_\alpha(y)$ and $\psi_\alpha(\phi)$ parts the following equations
\begin{eqnarray}
\hbar^{2} \partial^{2}_{y}\chi_\alpha(y) + (\alpha^{2}- 3\lambda_{s}^{2}\kappa e^{-2y /\sqrt{3}})\chi_\alpha(y) = 0, \\
\hbar^{2} \partial^{2}_{\phi}\psi_\alpha(\phi) + \left(\alpha^{2}-\frac{\lambda_{s}^{2} q^{2}}{4}e^{-2 \phi}\right)\psi_\alpha(\phi) =0
\end{eqnarray}
where $\alpha$ is the separation constant. \\
We consider the case for $\kappa > 0$, for which the general solutions to both parts are given by
\begin{eqnarray}
\chi_\alpha(y)&=&A_{1}I_{i\frac{\sqrt{3}\alpha}{\hbar}}(3\lambda_s\sqrt{\kappa}e^{-\frac{y}{\sqrt{3}}}/\hbar)+
A_{2}K_{i\frac{\sqrt{3}\alpha}{\hbar}}(3\lambda_s\sqrt{\kappa}e^{-\frac{y}{\sqrt{3}}}/{\hbar}), \\
\psi_\alpha(\phi)&=&B_{1}I_{i\frac{\alpha}{\hbar}}(\lambda_s q e^{-\phi}/2\hbar)+
B_{2}K_{i \frac{\alpha}{\hbar}}(\lambda_s q e^{-\phi}/{2\hbar}).
\end{eqnarray}
To avoid an infinite value of the wave function as $y\rightarrow - \infty$ and $\phi \rightarrow - \infty$ we must choose $\chi_\alpha(y)\sim K_{i\frac{\sqrt{3}\alpha}{\hbar}}(3\lambda_s \sqrt{\kappa}e^{-\frac{y}{\sqrt{3}}}/\hbar)$  and $\psi_\alpha(\phi) \sim K_{i \frac{\alpha}{\hbar}}(\lambda_s q e^{-\phi}/{2\hbar})$, which corresponds to the pre-big-bang regime for $\alpha^{2}<3 \lambda_s^{2}\kappa e^{-2y/\sqrt{3}}$ and where $K_{i\nu}(x)$ denotes the MacDonald function of imaginary order.
Using the result given in \cite{radoslaw} we normalize the $\chi_\alpha(y)$ part of the wave function in the following form
\begin{equation}
\int_{-\infty} ^{\infty} dy  \chi_{\alpha} ^*(y)\chi_{\alpha'}(y)= \delta(\alpha^2 -\alpha '^2),
\end{equation}
and a similar expression can be found for the $\phi$ part.

In order to write the WDWM equation (\ref{MWDW}) it is very useful to employ the next relation
\begin{equation}
f(x,p) \star g(x,p) = f\left( x + \frac{i\hbar}{2}{\buildrel{\rightarrow}\over{\partial}}_p, p - \frac{i\hbar}{2}{\buildrel{\rightarrow}\over{\partial}}_x\right)g(x,p),
\label{shift}
\end{equation}
from which we obtain the following equation
\footnotesize
\begin{equation}
\left( - \left( P_{y} - \frac{i\hbar}{2}{\buildrel{\rightarrow}\over{\partial}}_y\right)^2  + \left( P_{\phi} - \frac{i\hbar}{2}{\buildrel{\rightarrow}\over{\partial}}_\phi\right)^2 + \frac{1}{4}\lambda_{s}^{2}q^{2}  e^{-2\left(\phi + \frac{i\hbar}{2}{\buildrel{\rightarrow}\over{\partial}}_{P_{\phi}}\right)} -3\lambda_{s}^{2}\kappa e^{-\frac{2}{\sqrt{3}}\left(y + \frac{i\hbar}{2}{\buildrel{\rightarrow}\over{\partial}}_{P_{y}}\right)}   \right) \rho (y,P_{y},\phi, P_{\phi}) = 0.
\end{equation}
\normalsize
The last expression can be split into two equations corresponding to its real part
\footnotesize
\begin{equation}
\left[ -P_{y}^{2} + \frac{\hbar^{2}}{4}\partial^{2}_{y} + P_{\phi}^{2} - \frac{\hbar^{2}}{4}\partial^{2}_{\phi} + \frac{1}{4}\lambda_{s}^{2}q^{2} e^{-2 \phi}\cos \left(\hbar\partial_{P_{\phi}}\right) -3\lambda_{s}^{2}\kappa e^{-\frac{2}{\sqrt{3}}y} \cos\left( \frac{\hbar}{\sqrt{3}}\partial_{P_{y}} \right)  \right]\rho (y,P_{y},\phi, P_{\phi})= 0,
\label{real}
\end{equation}
\normalsize
and its imaginary part
\footnotesize
\begin{equation}
\left[ \hbar(P_y \partial_y) - \hbar(P_\phi \partial_\phi) -\frac{1}{4}\lambda_{s}^{2}q^{2} e^{-2\phi} \sin\left(\hbar \partial_{P_\phi}\right) + 3\lambda_{s}^{2}\kappa e^{-\frac{2}{\sqrt{3}}y} \sin\left(\frac{\hbar}{\sqrt{3}} \partial_{P_y}\right) \right] \rho(y,P_{y},\phi, P_{\phi}) = 0.
\label{imaginary}
\end{equation}
\normalsize
We propose $\rho(y,P_{y},\phi, P_{\phi}) = \rho_y(y, P_y) \rho_\phi(\phi, P_{\phi})$, and taking into account that $e^{i a \partial_x}f(x) = f(x+ ia)$ then from the two previous equations we obtain the following results: \\
For the function $\rho_y(y, P_y)$
\footnotesize
{\setlength\arraycolsep{-0.8em}
\begin{eqnarray}
\nonumber
&&\left[-P_{y}^2+\mu^2+\frac{\nu^4}{4P_{y}^2}\right]\rho_y(y,P_{y})+\frac{\hbar}{4P_{y}}\left\{ \left(\frac{\nu^2 u(y)}{i\hbar P_{y}}-\frac{2i u(y)}{\sqrt{3}}\right)
\left(\rho_y\left(y,P_{y}+\frac{i\hbar}{\sqrt{3}}\right)-\rho_y\left(y,P_{y}-\frac{i\hbar}{\sqrt{3}}
\right)\right) \right. \\ \nonumber
&&- \frac{u^{2}(y)}{\hbar} \left[\frac{1}{P_{y}+\frac{i\hbar}{\sqrt{3}}}
\left(\rho_y\left(y,P_{y}+\frac{2i\hbar}{\sqrt{3}}\right)-\rho_y(y,P_{y})\right)-\frac{1}{P_{y}-\frac{i\hbar}{\sqrt{3}}}
\left(\rho_y(y,P_{y})-\rho_y\left(y,P_{y}-\frac{2i\hbar}{\sqrt{3}}\right)\right)\right]  \\
&&\left. + \frac{\nu^2 u(y)}{i\hbar}\left[\frac{\rho_y(y,P_{y}+\frac{i\hbar}{\sqrt{3}})}
{P_{y}+\frac{i\hbar}{\sqrt{3}}}-\frac{\rho_y(y,P_{y}-\frac{i\hbar}{\sqrt{3}})}{P_{y}-\frac{i\hbar}{\sqrt{3}}}\right] \right \}- u(y)\left(\rho_y\left(y,P_{y}+\frac{i\hbar}{\sqrt{3}}\right)+
\rho_y\left(y,P_{y}-\frac{i\hbar}{\sqrt{3}}\right)\right)=0,
\label{ecndiferenciasy}
\end{eqnarray}
}
\normalsize

\begin{figure}
\begin{minipage}[t]{8cm}
\begin{center}
\includegraphics[scale=0.7]{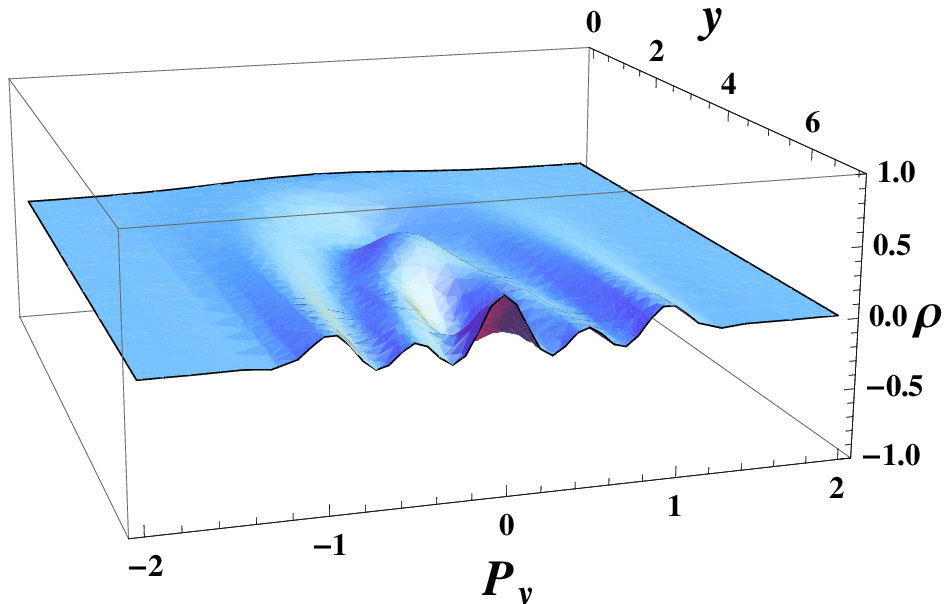}
\end{center}
\end{minipage}
\hfill
\begin{minipage}[t]{8cm}
\begin{center}
\includegraphics[scale=0.63]{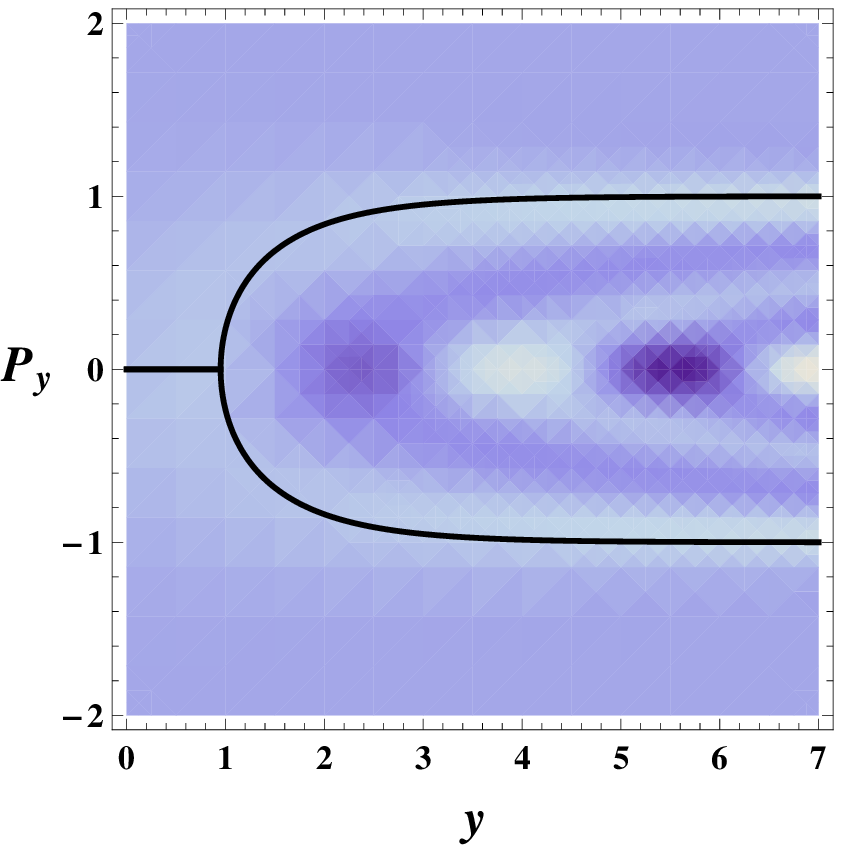}
\end{center}
\end{minipage}
\caption{{\footnotesize The Wigner function and its density are plotted in the $y$ variable ($\hbar=1, \alpha=1$). The figure on the left shows many oscillations due to the interference among wave functions of expanding and contracting universes. The thick curve on the right figure corresponds to the classical trajectory which coincides with the maximum of the Wigner function.}}
\label{fig:WFY1}
\end{figure}

\noindent and for the function $\rho_\phi(\phi, P_{\phi})$
\footnotesize
{\setlength\arraycolsep{-0.5em}
\begin{eqnarray}
\nonumber
&&\left[P_{\phi}^2-\mu^2-\frac{\nu^4}{4P_{\phi}^2}\right]\rho_\phi(\phi,P_{\phi})-\frac{\hbar}{4P_{\phi}} \left\{ \left(\frac{\nu^2 w(\phi)}{i\hbar P_{\phi}}
-2i w(\phi)\right)\left(\rho_\phi(\phi,P_{\phi}+i\hbar)-\rho_\phi(\phi,P_{\phi}-i\hbar)\right) \right. \\ \nonumber
&&-\frac{w^{2}(\phi)}{\hbar}\left[\frac{1}{P_{\phi}+i\hbar}(\rho_\phi(\phi,P_{\phi}+2i\hbar)-\rho_\phi(\phi,P_{\phi}))-
\frac{1}{P_{\phi}-i\hbar}(\rho_\phi(\phi,P_{\phi})-\rho_\phi(\phi,P_{\phi}-2i\hbar))\right] \\
&& \left. + \frac{\nu^2w(\phi)}{i\hbar}\left[\frac{\rho_\phi(\phi,P_{\phi}+i\hbar)}{P_{\phi}+i\hbar}-
\frac{\rho_\phi(\phi,P_{\phi}-i\hbar)}{P_{\phi}-i\hbar}\right]\right\}+w(\phi)(\rho_\phi(\phi,P_{\phi}+i\hbar)+
\rho_\phi(\phi,P_{\phi}-i\hbar))=0,
\label{ecndiferenciasphi}
\end{eqnarray}
}
\normalsize
\hspace{-0.3 cm} where we have defined the functions $u(y)=\frac{3\lambda_s^2\kappa e^{-\frac{2}{\sqrt{3}}y}}{2}$ and $w(\phi)=\frac{\lambda_s^2 q^{2}e^{-2\phi}}{8}$. \\
It is hard to solve the last two equations, so to obtain their solutions we will follow a different approach and will use the integral representation for the Wigner function. We can found its $y$ part by computing
{\setlength\arraycolsep{-0.4em}
\begin{eqnarray}
&&\rho_{y}(y,P_{y})=\frac{1}{2\pi}\int_{-\infty}^{\infty} \psi^*(y-\hbar q/2)\exp(-iqP_{y})\psi(y+\hbar q/2)dq \nonumber \\
&&=\frac{|A_{2}|^{2}}{2\pi}  \int_{-\infty}^{\infty} K_{i\frac{\sqrt{3}\alpha}{\hbar}}^*(3\lambda_s\sqrt{\kappa}e^{-\frac{(y-\hbar q/2)}{\sqrt{3}}}/\hbar)\exp(-iqP_{y})K_{i\frac{\sqrt{3}\alpha}{\hbar}}(3\lambda_s\sqrt{\kappa}e^{-\frac{(y+\hbar q/2)}{\sqrt{3}}}/\hbar) dq.
\end{eqnarray}
}
Defining the variables $\omega=e^{\hbar q/2\sqrt{3}}$ and $z=a=3\frac{\lambda_s}{\hbar}\sqrt{\kappa}e^{-y/\sqrt{3}}$, we get:
\begin{eqnarray}
\rho_{y}(y,P_{y})&=&\frac{\sqrt{3}|A_{2}|^{2}}{\pi \hbar}  \int_{0}^{\infty} K_{i\frac{\sqrt{3}\alpha}{\hbar}}(z\omega)\omega^{\sigma-\frac{1}{2}}K_{i\frac{\sqrt{3}\alpha}{\hbar}}(z/\omega)d\omega,
\end{eqnarray}
where $\sigma=-\frac{1}{2}-\frac{2\sqrt{3}}{\hbar}iP_{y}$ and $|A_{2}|^{2} = \frac{\sqrt{3}\sinh(\frac{\pi\sqrt{3}\alpha}{\hbar})}{\pi^{2}\hbar^{2}}$.

\begin{figure}
\begin{minipage}[t]{8cm}
\begin{center}
\includegraphics[scale=0.65]{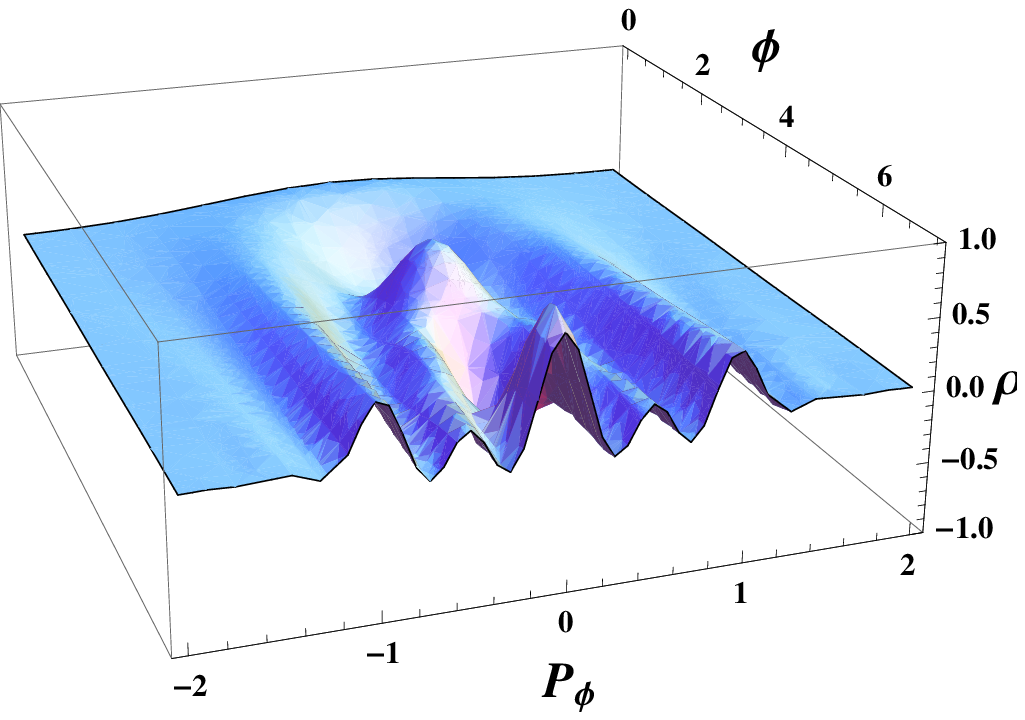}
\end{center}
\end{minipage}
\hfill
\begin{minipage}[t]{8cm}
\begin{center}
\includegraphics[scale=0.63]{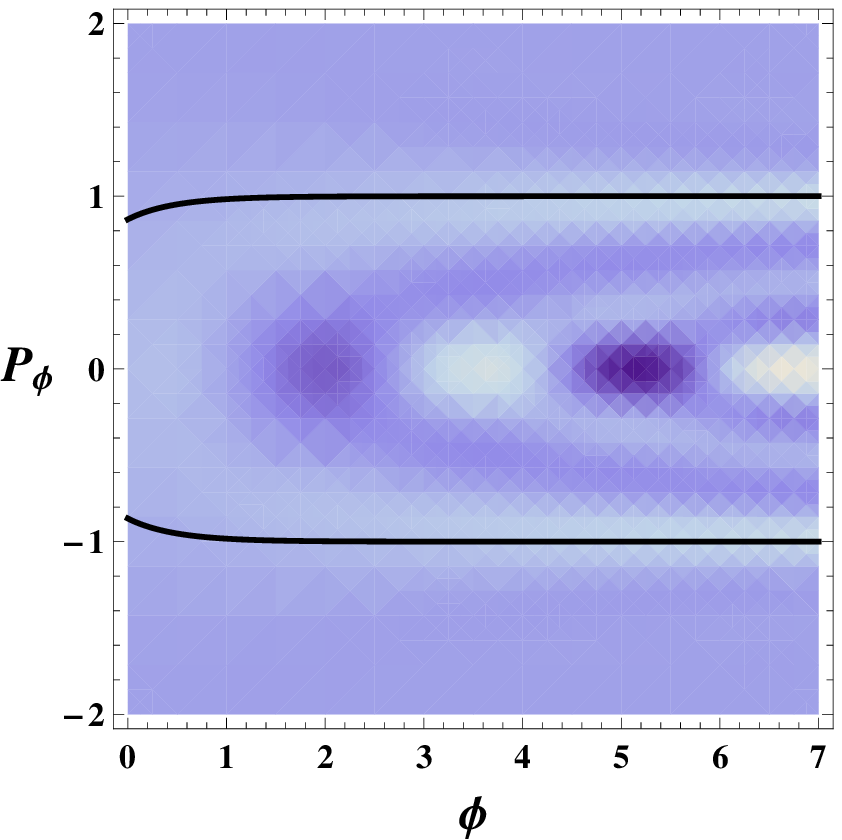}
\end{center}
\end{minipage}
\caption{{\footnotesize The Wigner function and its density are plotted in the $\phi$ variable ($\hbar=1, \alpha=1$). The figure on the left shows oscillations of higher amplitude with respect to the $y$ part. In the right figure the thick curve is the classical trajectory  and corresponds the maximum of the Wigner function (the $\phi < 0$ region is classical forbidden).}}
\label{fig:WFphi1}
\end{figure}

Using now the following result (see Sec. 19.6 formula (25) in \cite{ErdelyiTables} and the comment in \cite{CFZ})
\footnotesize
\begin{equation}
\int_0 ^{\infty} dw(wz)^{1/2} w^{\sigma -1} K_\mu (a/w) K_\nu (wz) = 2^{-\sigma -5/2} a^\sigma G^{40}_{04}\left(\frac{a^2 z^2}{16} {\bigg |} \frac{\mu - \sigma}{2}, \frac{-\mu - \sigma}{2}, \frac{1}{4}+ \frac{\nu}{2}, \frac{1}{4}- \frac{\nu}{2}\right),
\end{equation}
\normalsize
where $G^{40}_{04}\left(\frac{a^2 z^2}{16} {\bigg |} \frac{\mu - \sigma}{2}, \frac{-\mu - \sigma}{2}, \frac{1}{4}+ \frac{\nu}{2}, \frac{1}{4}- \frac{\nu}{2}\right)$ is a special case of Meijer's $G$ function (see Sec. 5.3 in \cite{Erdelyi})
\begin{equation}
 G^{mn} _{pq} \left( z \bigg | \begin{array}{lc}
a_{i}, & i=1,...,p \\
b_{j}, & j=1,...,q
\end{array}
\right),
\end{equation}
we obtain the following expression
\footnotesize
{\setlength\arraycolsep{-1.5em}
\begin{eqnarray}
\nonumber
&&\rho_{y}(y,P_{y})= \frac{\sinh(\frac{\pi\sqrt{3}\alpha}{\hbar})}{4 \pi^{3}\hbar^{2}} \frac{e^\frac{y}{\sqrt{3}}}{\lambda_s \sqrt{\kappa}}\left(\frac{3u(y)}{2\hbar^{2}}\right)^{-\frac{ \sqrt{3}}{\hbar}i P_{y}}\\
&&\times G_{04}^{40}\left(\frac{9 u^{2}(y)}{4 \hbar^{4}} \bigg | \frac{1}{4} + \frac{i\sqrt{3}}{\hbar}\left(\frac{\alpha}{2}+P_{y}\right),\frac{1}{4} + \frac{i\sqrt{3}}{\hbar}\left(\frac{-\alpha}{2}+P_{y}\right),\frac{1}{4}+\frac{i\sqrt{3}\alpha}{2\hbar},\frac{1}{4}-
\frac{i\sqrt{3}\alpha}{2\hbar}\right).
\label{eqn:WignerY}
\end{eqnarray}
}
\normalsize
Now, employing the Meijer's function property
\begin{equation}
 x^{\sigma}G^{mn}_{pq}\left( x \bigg | \begin{array}{lc}
a_{i}, & i=1,...,p \\
b_{j}, & j=1,...,q
\end{array}
\right) = G^{mn}_{pq}\left( x \bigg | \begin{array}{lc}
a_{i} +\sigma, & i=1,...,p \\
b_{j}+ \sigma, & j=1,...,q
\end{array}
\right),
\end{equation}
equation (\ref{eqn:WignerY}) can be written as
\footnotesize
{\setlength\arraycolsep{-0.5em}
\begin{eqnarray}
\nonumber
&&\rho_{y}(y,P_{y})= \frac{\sinh(\frac{\pi\sqrt{3}\alpha}{\hbar})}{4 \pi^{3}\hbar^{2}} \frac{e^\frac{y}{\sqrt{3}}}{\lambda_s \sqrt{\kappa}}\\
&&\times G_{04}^{40} \left( \frac{9 u^{2}(y)}{4 \hbar^{4}} \bigg |\frac{1}{4}+\frac{i \sqrt{3}}{2\hbar}(\alpha + P_{y}),\frac{1}{4}+\frac{i \sqrt{3}}{2\hbar}(-\alpha + P_{y}),\frac{1}{4}+\frac{i \sqrt{3}}{2\hbar}(\alpha - P_{y}),\frac{1}{4} + \frac{i \sqrt{3}}{2\hbar}(-\alpha - P_{y}) \right).
\end{eqnarray}
}
\normalsize
\hspace{-0.45em} It is possible to verify that this Wigner function indeed satisfy equation (\ref{ecndiferenciasy}). \\
Performing a similar procedure we can obtain the following Wigner function for the $\phi$ part
\footnotesize
{\setlength\arraycolsep{-1em}
\begin{eqnarray}
\nonumber
&&\rho_{\phi}(\phi,P_{\phi})= \frac{\sinh(\frac{\pi\alpha}{\hbar})}{2 \pi^{3}\hbar^{2}} \frac{e^\phi}{\lambda_s q}\\
&&\times G_{04}^{40} \left( \frac{w^{2}(\phi)}{4 \hbar^{4}} \bigg |\frac{1}{4}+\frac{i}{2\hbar}(\alpha + P_{\phi}),\frac{1}{4}+\frac{i}{2\hbar}(-\alpha + P_{\phi}),\frac{1}{4}+\frac{i}{2\hbar}(\alpha - P_{\phi}),\frac{1}{4} + \frac{i }{2\hbar}(-\alpha - P_{\phi}) \right),
\end{eqnarray}
}
\normalsize
which fulfills Eqn. (\ref{ecndiferenciasphi}).

We can gain some physical insight if we plot the Wigner functions of the corresponding $y$ and dilaton parts for several values of $\alpha$. For the $y$ part and $\alpha=1$ Fig. \ref{fig:WFY1} shows that the classical trajectory is near the highest peaks of the Wigner function. For values of $\alpha$ smaller than one there are less oscillations but the classical trajectory does not correspond to the highest peaks, in fact, there is an ample region where the Wigner function is large. For values of $\alpha$ bigger than one it can be observed an increment in the number of oscillations of Wigner function and the peaks of the oscillations are far away from the classical trajectory (these plots are not showed in the paper). We conclude that the quantum interference effects are enhanced for larger values of $\alpha$. A similar behavior is obtained for the dilaton part for $\alpha$, nevertheless from Fig. \ref{fig:WFphi1} its amplitude is larger. The parameter $\alpha$ can be interpreted as the energy of the $y$ and $\phi$ parts, the matter (dilaton axion part) has positive energy and any fluctuation of it induces a variation of the energy in the gravitational part (the scale factor) in order to vanish the total energy.

\section{Final Remarks}
In this paper we have presented the WWGM formalism for a gravitational field coupled
to matter in the flat superspace (and flat phase superspace) where the Stratonovich-Weyl
quantizer, the star product and the Wigner functional are obtained.
These results can be used in general situations but in a first
approach we applied \cite{Cordero:2011xa} the formalism to some
interesting minisuperspace models widely studied in the literature,
in particular we studied here the axion-dilaton quantum cosmology
model with non-vanishing spatial curvature. We found the corresponding
WDWM equation and the equivalent differential-difference equation. These
equations have an exact solution in terms of the Meijer's
functions. We have seen that the parameter $\alpha$ has an
interpretation of energy for the components $y$ and $\phi$. For an
arbitrary total constant energy any fluctuation in the energy of
matter does induce a corresponding fluctuation in the gravitational
part in such a way that both are compensated. Moreover, we found
that for small values of $\alpha$ there are fewer oscillations in
the gravitational part of the Wigner function but the classical trajectory
does not correspond to the highest peaks. In the case of bigger values of $\alpha$ there
is a clear increment in the number of oscillations of Wigner
function, being its peaks far away from the classical trajectory. We
observe also that the quantum interference effects for larger values
of $\alpha$ are enhanced. The same situation is regarded for the
dilaton part but with larger amplitude oscillations. It is worth
to mention that the construction presented here can be employed to
treat other cosmological models.

\vskip 2truecm

\centerline{\bf Acknowledgments}

\vskip 1truecm

The work of R. C., H. G.-C. and F. J. T. was partially supported by SNI-M\'exico, CONACyT research grants: J1-60621-I, 103478 and 128761. In addition R. C. and F. J. T. were partially supported by COFAA-IPN and by SIP-IPN grants 20111070 and 20110968.

\end{document}